\documentclass[12pt]{article}
\usepackage[round]{natbib}
\usepackage[english]{babel}
\usepackage[utf8]{inputenc}
\usepackage{graphicx}
\usepackage{rotating}
\usepackage{pdflscape}
\usepackage{float}
\usepackage[bottom]{footmisc}
\usepackage{comment}
\usepackage{mathtools}
\usepackage{extpfeil}
\usepackage{dsfont}
\usepackage{physics}
\usepackage{bm}
\usepackage{url}
\usepackage{subcaption}
\usepackage{booktabs}

\usepackage{amsmath}

\usepackage{epsfig}
\usepackage{setspace} 
\usepackage{array}
\usepackage{multirow}
\usepackage[dvipsnames]{xcolor}
\usepackage{subcaption}

\newcolumntype{L}[1]{>{\raggedright\let\newline\\\arraybackslash\hspace{0pt}}m{#1}}
\newcolumntype{C}[1]{>{\centering\let\newline\\\arraybackslash\hspace{0pt}}m{#1}}
\newcolumntype{R}[1]{>{\raggedleft\let\newline\\\arraybackslash\hspace{0pt}}m{#1}}

\begin{document}

\title{A more interpretable regression model for count data with excess of zeros}

\author{Gustavo H. A. Pereira$^{1}$ \and Jeremias Leão$^2$ \and Manoel Santos-Neto $^3$ \and Jianwen Cai$^4$}
\date{}

\maketitle

\footnotetext[1]{Department of Statistics, Federal University of São Carlos, Rod. Washington Luís, km 235 - SP-310 - São Carlos, CEP 13565-905,  Brazil. Email: gpereira@ufscar.br}
\footnotetext[2]{Department of Statistics, Federal University of Amazonas, Brazil}
\footnotetext[3]{Department of Statistics, Federal University of Ceará, Brazil}
\footnotetext[4]{Department of Biostatistics, Gillings School of Global Public Health, University of North Carolina at Chapel Hill, Chapel Hill, NC, USA.}


\begin{abstract}
\noindent Count data are common in medical research. When these data have more zeros than expected by the most used count distributions, it is common to employ a zero-inflated regression model. However, the interpretability of these models is much lower than the most used count regression models. In this work, we introduce a more interpretable regression model for count data with excess of zeros based on a reparameterization of the zero-inflated Poisson distribution.  We discuss inferential and diagnostic tools and perform a Monte Carlo simulation study to evaluate the performance of the maximum likelihood estimator.  Finally, the usefulness of the proposed regression model is illustrated through an application on children mortality. 
\end{abstract}

\vspace{0mm} \noindent {\textbf{Keywords}}: Count data, quantile residual, zero-inflated data, zero-inflated Poisson regression model.


\section{Introduction}
\label{sec:introd}


Count data are common in several areas of medicine such as immunology \citep{brazel2024baseline}, cardiology \citep{doan2024identifying}, urology \citep{hutchison2024comparison}, pediatrics \citep{stordal2024genotypes}, psychiatry \citep{byhoff2024mental} and neurology \citep{joundi2024association}. 
A possible way to study the relationship between these variables and a set of explanatory variables is assuming that the response has a certain count probability distribution such as Poisson, negative binomial, Poisson inverse Gaussian, among many others. However, it is not unusual that the response variable has more zeros than expected by the best known count distributions.

When there is an excess of zeros, the response variable is usually modeled using a zero-inflated regression model. These models assume that the response variable is a mixture of a degenerated distribution at zero and a known count probability distribution. Several zero-inflated regression models were proposed considering as the count probability distribution the Poisson distribution \citep{lambert1992zero}, the negative binomial distribution \citep{ridout2001score}, the logarithmic distribution \citep[page 492]{rigby2019distributions}, among others.

In the zero-inflated regression models, there are two means related to the response variable. The first is the mean of the response variable. The other is the mean of the count probability distribution that composes the distribution of the response variable. We usually want to model the former mean as a function of covariates. However, in almost all of the zero-inflated regression models, the latter mean is modeled. The reason is that the vector of parameters of the distribution of the response variable in these models includes the mean of the count probability distribution and not the mean of the response variable. 


To the best of our knowledge, there is only one zero-inflated regression model in which the mean of the response variable is directly modeled. This model is based on a reparameterization of the zero-inflated Poisson distribution and was proposed by \citet{long2014marginalized}. This model can be fitted in the gamlss package \citep{stasinopoulos2008generalized} of software R and was considered for example by \citet{seidel2020predicting} and \citet{sims2024application}. However, this parameterization does not change the other parameter of the distribution. As a result, the variance of the distribution is not a simple function of the model parameters and one of the parameters is not interpretable in this parameterization in most of the practical situations.

This work proposes a zero-inflated regression model based on a novel reparameterization of the zero-inflated Poisson distribution. This parameterization is more useful than the existing ones because one of the parameters of the distribution is the mean and the other a dispersion parameter. As a result, the proposed model is more interpretable than the other zero-inflated regression models. We assume that both  parameters of the distribution of the response are  functions of covariates, and so the proposed model has a structure similar to a double generalized linear model \citep{smyth1989generalized}.

The remainder of this paper is organized as follows. Section \ref{sec:param} proposes a reparameterization of the zero-inflated Poisson distribution. Section \ref{sec:model} introduces the regression model associated with this novel reparameterization. Diagnostic tools for this regression model are discussed in Section \ref{sec:diagn}. In Section \ref{sec:simul}, Monte Carlo simulation studies are performed to evaluate the performance of the maximum likelihood estimators of the parameters of the proposed regression model. The usefulness of our model is illustrated through an application on children mortality presented in Section \ref{sec:appl}.  Concluding remarks are provided in Section \ref{sec:conc}.


\section{Parameterizations of the ZIP}
\label{sec:param}

The first version of the zero-inflated Poisson (ZIP) distribution was introduced by \cite{lambert1992zero}. In this setup, the probability mass function is given by:
\begin{equation}
\label{eq:zip1}
\Pr(Y=y | \lambda,p)=
\begin{cases}
p + (1-p)e^{-\lambda}, &\text{if $y=0$}, \\
(1-p)e^{-\lambda}\lambda^y/y!,   &\text{if $y=1, 2, 3, \dots,$}
\end{cases}
\end{equation}
where $\lambda > 0$, $0 < p< 1$. In this parameterization, denoted by ZIP1, $\text{E}(Y) = (1-p)\lambda$ and $\text{Var}(Y)=\lambda(1-p)(1+\lambda p)$. 
This probability mass function is derived based on the assumption that data are composed of two (unobservable) subpopulations.  The response  variable is zero for the first subpopulation and is Poisson distributed for the other. The parameter $p$ is the probability of belonging to the first subpopulation and $\lambda$ is the mean of the response variable for the second subpopulation. However, since the two subpopulations are not observable, the parameters $p$ and $\lambda$ are of less interest in practice.

A regression model is more easily interpretable when one the parameters of the distribution of the response variable is the mean or the median.  \cite{long2014marginalized} proposed a new parameterization of the ZIP distribution, in which one of the parameters is the mean.
From the ZIP1 parameterization, they considered that $\mu^\ast=(1-p)\lambda$ and $\delta^\ast=p$, i.e, $\lambda=\mu^\ast/(1-\delta^\ast)$ and $p=\delta^\ast$. Therefore they obtain from (\ref{eq:zip1}) the following probability mass function
\begin{equation*}
\label{eq:zip2}
\Pr(Y=y | \mu^\ast,\delta^\ast)=
\begin{cases}
\delta^\ast + (1-\delta^\ast)e^{-\left(\dfrac{\mu^\ast}{1-\delta^\ast}\right)}, &\text{if $y=0$}, \\
\dfrac{(\mu^\ast)^y}{y!(1-\delta^\ast)^{y-1}}
e^{-\left(\dfrac{\mu^\ast}{1-\delta^\ast}\right)}, &\text{if $y=1, 2, 3, \dots,$}
\end{cases}
\end{equation*}
where $\mu^\ast > 0$, $0 < \delta^\ast < 1$. In this parameterization, $\text{E}(Y) = \mu^\ast$ and $\text{Var}(Y)=\mu^\ast[1 + \mu^\ast\delta^\ast/(1-\delta^\ast)]$. We will refer to this parameterization as ZIP2. Note that from the ZIP1 to the ZIP2 parameterization, the parameter $\delta^\ast$ was not changed. As a result, $\delta^\ast$ is not of direct interest and the variance is not a simple function of the parameters. 

Here we propose a novel parameterization of the ZIP distribution, in which both parameters are of direct interest. Moreover, in this parameterization, the variance of the ZIP distribution is a simple function of the parameters. The ZIP distribution in our proposed parameterization is indexed by the mean and a dispersion parameter. 
 From the ZIP1 parameterization, we consider that $\mu=(1-p)\lambda$ and $\phi=p\lambda$, i.e, $p=\phi/(\mu+\phi)$ and $\lambda=\mu+\phi$. Therefore we obtain from (\ref{eq:zip1}) the following probability mass function
\begin{equation}
\label{eq:zip3}
\Pr(Y=y \mid \mu,\phi)=
\begin{cases}
\dfrac{\phi + \mu e^{-(\mu+\phi)}}{\mu + \phi}, &\text{if $y=0$}, \\
\dfrac{\mu(\mu+\phi)^{y-1} e^{-(\mu+\phi)}}{y!}, &\text{if $y=1, 2, 3, \dots,$}
\end{cases}
\end{equation}
where $\mu > 0$, $\phi > 0$. In this parameterization, $\text{E}(Y) = \mu$ and $\text{Var}(Y)=\mu(1+\phi)$. We will refer to this parameterization as ZIP3. 

\section{A regression model}
\label{sec:model}

Considering the interpretability advantages of ZIP3 parameterization, it is very convenient to use when the response variable has a high proportion of zeros. We define in this section a regression model based on the ZIP3 parameterization and use it to fit real data in Section \ref{sec:appl}. However, before introducing our regression model, we obtain some results that enable us to use the model in an useful computational framework.

The expression \eqref{eq:zip3} can be rewritten as follows:
\begin{align}\label{eq:zip31}
\Pr(Y=y \mid \mu,\phi) = \left[\frac{\phi + \mu e^{-(\mu+\phi)}}{\mu + \phi}\right]^{I(y=0)} \left[\frac{\mu(\mu+\phi)^{y-1} e^{-(\mu+\phi)}}{y!}\right]^{(1-I(y=0)) }, \quad y \in \mathbb{Z}_0^{+},
\end{align}
where $I(\cdot)$ is an indicator function, i.e., $I(y=0) = 1$ if $y=0$ and 0 otherwise.
Equation \eqref{eq:zip31} enables us to obtain the logarithm of the probability function. The expression can be stated as follows:
\begin{align}\label{eq:zip32}
\ell(\mu, \phi \mid y) &=   I(y=0)\left[\log(\phi + \mu e^{-(\mu+\phi)}) - \log(\mu + \phi)\right] + (1 - I(y=0))\left[\log(\mu) + (y-1) \right. \notag \\ 
&\left. \quad \log(\mu + \phi) - \mu - \phi - \log(y!)\right].
\end{align}

Additionally, the partial derivatives of first and second order of Equation \eqref{eq:zip32} with respect to the parameters indexing the ZIP3 distribution are given by:
\begin{align}\label{eq:zip33}
\partial_\mu \ell(\mu, \phi \mid y) &= I(y=0) \hspace{-0.1cm} \left[\frac{(1-\mu)}{(\mu + \phi \,\textrm{e}^{\mu + \phi})} - \frac{1}{(\mu+\phi)}\right] + (1 - I(y=0)) \hspace{-0.1cm} \left[\frac{1}{\mu} + \frac{(y-1)}{(\mu + \phi)} - 1\right], \notag\\
\partial_{\mu\mu}^2 \ell(\mu, \phi \mid y) &=  I(y=0) \hspace{-0.1cm} \left[\frac{\phi(\mu-2)\textrm{e}^{\mu + \phi} -1}{(\mu + \phi \,\textrm{e}^{\mu + \phi})^2} + \frac{1}{(\mu+\phi)^2}\right] \hspace{-0.05cm} + \hspace{-0.05cm} (1 - I(y=0)) \hspace{-0.1cm} \left[-\frac{1}{\mu^2} - \frac{(y-1)}{(\mu + \phi)^2}\right],\notag\\ 
\partial_\phi \ell(\mu, \phi \mid y) &= I(y=0) \hspace{-0.1cm} \left[\frac{\textrm{e}^{\mu + \phi}- \mu}{(\mu + \phi \,\textrm{e}^{\mu + \phi})} - \frac{1}{(\mu+\phi)}\right] + (1 - I(y=0)) \hspace{-0.1cm} \left[\frac{(y-1)}{(\mu + \phi)} -1\right],\\
\partial_{\phi\phi}^2 \ell(\mu, \phi \mid y) &= I(y=0) \hspace{-0.1cm} \left[\frac{\textrm{e}^{\mu + \phi}[\mu(\phi + 2) - \textrm{e}^{\mu + \phi}]}{(\mu + \phi \,\textrm{e}^{\mu + \phi})^2} + \frac{1}{(\mu+\phi)^2}\right] + (1 - I(y=0)) \hspace{-0.1cm} \left[- \frac{(y-1)}{(\mu + \phi)^2}\right],\notag\\ 
\partial_{\mu\phi}^2\ell(\mu, \phi \mid y) &=I(y=0) \hspace{-0.1cm} \left[\frac{(\mu - 1)(\phi + 1)\textrm{e}^{\mu + \phi}}{(\mu + \phi \,\textrm{e}^{\mu + \phi})^2} + \frac{1}{(\mu+\phi)^2}\right] + (1 - I(y=0)) \hspace{-0.1cm} \left[- \frac{(y-1)}{(\mu + \phi)^2}\right]. \notag
\end{align}

\subsection{The ZIP3 regression model}
\label{sec:model2}

In ZIP3 regression, the response variables 
\(\mathbf{Y} = (Y_1, \dots, Y_n)^\top\) are independent and follow a ZIP distribution with parameters \(\mu_i\) and \(\phi_i\) as defined in the ZIP3 parameterization presented in (\ref{eq:zip3}). Moreover, the parameters  \(\boldsymbol{\mu} = (\mu_1, \dots, \mu_i, \dots, \mu_n)^\top\) and \(\boldsymbol{\phi} = (\phi_1, \dots, \phi_i, \dots, \phi_n)^\top\) satisfy
\begin{equation}\label{eq:reg}
 g_1(\mu_i) = \mathbf{x}_i^\top \boldsymbol{\beta} = \eta_{i}, \quad \text{and} \quad  g_2(\phi_i) = \mathbf{z}_i^\top\boldsymbol{\gamma} = \varsigma_i,
\end{equation}
for (vectors of) covariates  \(\mathbf{x}_i^\top= (1, x_{i2}, \dots, x_{iq_1})\), and \(\mathbf{z}_i^\top = (1, z_{i2}, \dots, z_{iq_2})\) with \(\boldsymbol{\beta} = (\beta_1, \dots, \beta_{q_{1}})^\top\) and \(\boldsymbol{\gamma} = (\gamma_1, \dots, \gamma_{q_{2}})^\top\) being the parameter vectors associated with $\mathbf{x}_i$ and $\mathbf{z}_i$, respectively. Additionally, \(g_k(\cdot),  (k = 1, 2)\) specifies the link between the random and the systematic components and is a strictly monotonic and twice differentiable function. It follows that the ZIP3 regression log-likelihood is:
\begin{align}\label{eq:zip7}
\ell_1(\bm \mu, \bm \phi \mid \bm y) &= \sum_{i = 1}^n \ell(\mu_i, \phi_i \mid  y_i) \notag \\ 
&=\sum_{i = 1}^n \varrho_i\left[\log(\phi_i + \mu_i e^{-(\mu_i+\phi_i)}) - \log(\mu_i + \phi_i)\right] + (1 - \varrho_i)\left[\log(\mu_i) + (y_i-1) \right. \notag \\ 
&\left. \quad \log(\mu_i + \phi_i) - \mu_i - \phi_i - \log(y_i!)\right].
\end{align}
where \(\varrho_i = I(y_i=0)\), \(\mu_i = g_1^{-1}(\mathbf{x}_i^\top \boldsymbol{\beta})\), and \(\phi_i = g_2^{-1}(\mathbf{z}_i^\top \boldsymbol{\gamma})\).

Let \(\boldsymbol{\theta} = (\boldsymbol{\beta}^\top, \boldsymbol{\gamma}^\top)^\top\) be the unknown $s$-dimensional \((s := q_1 + q_2)\) parameter in models \eqref{eq:reg}. The maximum likelihood (ML) estimator \(\widehat{\boldsymbol{\theta}} = (\widehat{\boldsymbol{\beta}}^\top, \widehat{\boldsymbol{\gamma}}^\top)^\top\) of \(\boldsymbol{\theta}\) is the solution of the \(s\)-dimensional score equation
\begin{align}
U(\widehat{\boldsymbol{\theta}}) = \boldsymbol{0},   
\end{align}
where \(U(\boldsymbol{\theta}) =  (U_{\boldsymbol{\beta}}(\boldsymbol{\theta})^\top, U_{\boldsymbol{\gamma}}(\boldsymbol{\theta})^\top)^\top\) is the score vector. The score function is given by taking the first derivative of the log-likelihood function, \eqref{eq:zip7}, with respect to each element of \(\boldsymbol{\theta}\). By the chain rule, it follows that

$$U_{\beta_j}(\boldsymbol{\theta}) = \sum\limits_{i=1}^n \pdv{\ell(\mu_i, \phi_i \mid  y_i)}{\mu_i}\pdv{\mu_i}{\eta_i}\pdv{\eta_i}{\beta_i} = \sum\limits_{i=1}^n d_{\mu_i} l_{\mu_i} x_{ij},$$

and 

$$U_{\gamma_j}(\boldsymbol{\theta}) = \sum\limits_{i=1}^n \pdv{\ell(\mu_i, \phi_i \mid  y_i)}{\phi_i}\pdv{\phi_i}{\varsigma_i}\pdv{\varsigma_i}{\gamma_i} = \sum\limits_{i=1}^n d_{\phi_i} l_{\phi_i} z_{ij},$$
where
\begin{align*}
d_{\mu_i} &= \varrho_i\left[\frac{(1-\mu_i)}{(\mu_i + \phi_i \,\textrm{e}^{\mu_i + \phi_i})} - \frac{1}{(\mu_i+\phi_i)}\right] + (1 - \varrho_i)\left[\frac{1}{\mu_i} + \frac{(y_i-1)}{(\mu_i + \phi_i)} - 1\right]; \\
l_{\mu_i} & = \frac{1}{g_1'(\mu_i)};\\
d_{\phi_i} &= \varrho_i\left[\frac{\textrm{e}^{\mu_i + \phi_i}- \mu_i}{(\mu_i + \phi_i \,\textrm{e}^{\mu_i + \phi_i})} - \frac{1}{(\mu_i+\phi_i)}\right] + (1 - \varrho_i)\left[\frac{(y_i-1)}{(\mu_i + \phi_i)} -1\right]; \\
l_{\phi_i} & = \frac{1}{g_2'(\phi_i)},
\end{align*}
and the score vector can be written compactly as
$$
U_{\boldsymbol{\beta}}(\boldsymbol{\theta}) = \mathbf{X}^\top \mathbf{L}_{\boldsymbol{\mu}} [(\mathbf{y}^\ast - \boldsymbol{\mu}^\ast) +\boldsymbol{\varrho}\odot\mathbf{c}_1] 
\quad 
\text{and}
\quad 
U_{\boldsymbol{\gamma}}(\boldsymbol{\theta}) = \mathbf{Z}^\top \mathbf{L}_{\boldsymbol{\phi}} [(\mathbf{y}^\ast - \mathbf{1}_n) +\boldsymbol{\varrho}\odot\mathbf{c}_2], 
$$
where \(\odot\) represent the Hadamard product \citep{r1974hadamard}, \(\mathbf{X}\) is a \(n \times q_1\) matrix with the \(i\)th row given by \(\mathbf{x}_i^\top\), \(\mathbf{Z}\) is a \(n \times q_2\) matrix with the \(i\)th row given by \(\mathbf{z}_i^\top\), \(\mathbf{L}_{\boldsymbol{\mu}} = \textrm{diag}(l_{\mu_1}, \dots, l_{\mu_n})\), \(\mathbf{L}_{\boldsymbol{\phi}} = \textrm{diag}(l_{\phi_1}, \dots, l_{\phi_n})\), \({\mathbf{y}^\ast}^\top = (\frac{y_1 - 1}{\mu_1 + \phi_1}, \dots, \frac{y_n - 1}{\mu_n + \phi_n})\),  \(\mathbf{1}_n^\top = (1, \dots, 1)\), \(\boldsymbol{\mu}^\ast = (1 - \frac{1}{\mu_1}, \dots, 1 - \frac{1}{\mu_n})\), \(\mathbf{c}_1^\top = (\frac{1-\mu_1}{\mu_1 + \phi_1 \,\textrm{e}^{\mu_1 + \phi_1}} + \frac{\mu_1 -1}{\mu_1} - \frac{y_1}{\mu_1 + \phi_1}), \dots, \frac{1-\mu_n}{\mu_n + \phi_n \,\textrm{e}^{\mu_n + \phi_n}} + \frac{\mu_n -1}{\mu_n} - \frac{y_n}{\mu_n + \phi_n}) \), \(\mathbf{c}_2^\top = (\frac{\textrm{e}^{\mu_1 + \phi_1}- \mu_1}{\mu_1 + \phi_1 \,\textrm{e}^{\mu_1 + \phi_1}} + \frac{\mu_1 + \phi_1 - y_1}{\mu_1 + \phi_1}, \dots, \frac{\textrm{e}^{\mu_n + \phi_n}- \mu_1}{\mu_n + \phi_n \,\textrm{e}^{\mu_n + \phi_n}}  +  \frac{\mu_n + \phi_n -y_n}{\mu_n + \phi_n}) \), and \(\boldsymbol{\varrho}^\top = (\varrho_1, \dots, \varrho_n)\).

With the results presented in Equation \eqref{eq:zip33}, we can readily integrate the ZIP3 distribution into the distribution family framework of the \verb|R| package \verb|gamlss|~\citep{gamlss}. To make this integration,
we developed a suite of functions, encompassing the pseudo-random number generator, the probability density function, the quantile function, and the cumulative distribution function.  All codes are available on \verb|Github| through the link  \url{https://github.com/statlab-oficial/ZIP3}. With this integration, one can fit the ZIP3 regression model taking advantage of all inferential and diagnostic tools of the gamlss package. We intend to include the ZIP3 regression model in the gamlss package after this work is published.


\section{Diagnostic analysis}
\label{sec:diagn}

We propose using a residual and a global influence measure for model diagnostics for ZIP3 regression model.

\subsection{Residual analysis}
\label{sec:resid}

When the response of a regression model is discrete, Pearson and deviance residuals are also discrete. As a result, these residuals have a considerable probability of not detecting lack of fit \citep{feng2020comparison}. For this reason, when the response is discrete, it is better to use the randomized quantile residual \citep{dunn1996randomized} to evaluate the goodness of fit of the regression model. For the ZIP3 regression model, the randomized quantile residual is given by 
\begin{equation}\label{eq:resid}
q_i = \Phi^{-1}(u_i),
\end{equation}
where $u_i$ is a uniform random variable on the interval $(F(y_i-1;\hat{\mu}_i,\hat{\phi}_i),
F(y_i;\hat{\mu}_i,\hat{\phi}_i))$, $\Phi(\cdot)$ and $F(\cdot)$ are the cumulative distribution function of the standard normal distribution and of the ZIP3 distribution, respectively, and $\hat{\mu}_i $ and $\hat{\phi}_i$ are the ML estimates of the parameters $\mu_i$ and $\phi_i$, respectively. The residual $q_i$ is asymptotically standard normally distributed under the correct model.

\subsection{Global influence}
\label{sec:influ}

According to \citet{dennis1988likelihood}, the likelihood displacement \citep[page 182]{cook1982residuals}  is the most useful measure for identifying influential observations. It has a similar expression for all parametric regression models and has been widely used in recent works \citep{cortes2023new,fabio2023diagnostic,
ibacache2023influence}. For the ZIP3 regression model, the likelihood displacement is given by
\begin{equation}\label{eq:ldi}
\text{LD}_i = 2[\ell_1(\bm \hat{\mu}, \bm \hat{\phi} \mid \bm y) - \ell_{1(i)}(\bm \hat{\mu}_{(i)}, \bm \hat{\phi}_{(i)} \mid \bm y_{(i)})],
\end{equation}
where $\ell_1(\bm \hat{\mu}, \bm \hat{\phi} \mid \bm y)$ and $\ell_{1(i)}(\bm \hat{\mu}_{(i)}, \bm \hat{\phi}_{(i)} \mid \bm y_{(i)})$ are the log-likelihood functions for the complete data and for data without the 
$i$th observation,
respectively. The calculation of $\text{LD}_i$ for the $n$ observations requires the estimation of $(n+1)$ ZIP3 regression model.  However, this is not an issue, since the fit of a ZIP3 regression model is fast using the code developed in this work.

\section{Simulation studies}
\label{sec:simul}

We conducted Monte Carlo (MC) simulation studies to evaluate the performance of the ML estimators of the ZIP3 regression model parameters using small and moderate sample sizes. Scenario 1 considers the following: sample sizes $n \in \{50, 100, 200, 500\}$ and values for the parameter as presented in (\ref{sim1_cen}), 
\begin{align}\label{sim1_cen}
 \log({\bm \mu}_{i}) =  -1.0 + 1.0 x_{i1} + 0.5 x_{i2}\qquad\mbox{and}\qquad\log(\phi_{i}) =  1.0 + 0.5 z_{i1},
\end{align}
where the covariates $x_{i1}$ and $z_{i1}$, for $i = 1,\ldots,n$ were generated from the standard uniform distribution and $x_{i2}$ was generated from the Bernoulli distribution with parameter $0.5$. The number of MC replications was 5,000 and all simulations were performed using the \texttt{R} programming language. 

For each value of the parameter and sample size, we report the bias (B) and mean squared error (MSE) of the ML estimators in Table  \ref{tab:1}. Note that, as the sample size increases, the bias and mean squared error of the ML estimators decrease, as expected. The biases are  small, except for $\hat{\beta}_0$, for which the bias is small only for $n \geq 100$. The mean squared errors are moderate when $n = 50$, but small for all ML estimators  when $n = 500$. 

\begin{table}[H]
\centering
\caption{Bias and mean square error of the ML estimator in Scenario 1.}
\renewcommand{\arraystretch}{1.3}
\resizebox{\linewidth}{!}{
\begin{tabular}{c crcccc| cccc}\hline
\multirow{2}*{$n$}& \multicolumn{6}{c|}{\textbf{Mean parameter}} & \multicolumn{4}{c}{\textbf{Precision parameter}} \\ \cline{2-11}
            &$\mbox{B}(\widehat{\beta}_{0})$ & $\mbox{B}(\widehat{\beta}_{1})$  &$\mbox{B}(\widehat{\beta}_{2})$  & $\mbox{MSE}(\widehat{\beta}_{0})$  & $\mbox{MSE}(\widehat{\beta}_{1})$ & $\mbox{MSE}\widehat{\beta}_{2})$ &
$\mbox{B}(\widehat{\phi}_{0})$ & $\mbox{B}(\widehat{\phi}_{1})$  & $\mbox{MSE}(\widehat{\phi}_{0})$  & $\mbox{MSE}(\widehat{\phi}_{1})$ \\ \hline 
50     & $-$0.2102 &    0.0370 & $-$0.0187 & 1.8188 & 1.9312 & 2.5304 & $-$0.0949 &    0.0354 & 0.4211 & 0.8702\\
100    & $-$0.0978 & $-$0.0019 &    0.0107 & 0.4479 & 0.8802 & 0.2858 & $-$0.0323 & $-$0.0049 & 0.1067 & 0.2855\\
200    & $-$0.0598 &    0.0155 &    0.0161 & 0.1999 & 0.3351 & 0.1292 & $-$0.0156 &    0.0042 & 0.0433 & 0.1078\\
500    & $-$0.0305 &    0.0105 &    0.0106 & 0.0734 & 0.1305 & 0.0489 & $-$0.0102 &    0.0057 & 0.0167 & 0.0411\\
\hline
\end{tabular}
}
\label{tab:1}
\end{table}

We also considered two other scenarios. In the first, from the Scenario 1, we changed the vector of parameters $\boldsymbol{\beta}$ and in the other we modified $\boldsymbol{\gamma}$. Results for these scenarios are similar to Scenario 1 and are not included here for brevity.

\section{Application to data on children mortality in Oromia - Ethiopia}
\label{sec:appl}

Children mortality is an important issue in Sub-Saharan Africa countries. For the year of 2022, it is estimated that $56.7\%$ of deaths in children under 5 years old in the world were in the Sub-Saharan Africa countries \citep{un2024}. Ethiopia has a large population and a high mortality rate for children under 5 years old (46 deaths per 1000 live births) and hence it is a country where there are a large number of deaths in children under 5 years old. 

To design policies and strategies to reduce the under-five mortality rate, it is valuable to identify covariates that are related to this rate. Here, we consider data about the region of Oromia, Ethiopia, collected by \citet{et2021} and used by \citet{argawu2023zero}. Data have information on 691 mothers from 15 to 49 years age and the response variable is the number of under-five children deaths. The following covariates are available:  mother’s age, place of residence (urban or rural), mother’s education level, literacy (can read or can not read), marital status, mother’s religion, source of water (improved or not improved), time to get water, types of toilet facility (improved or not improved), type of cooking fuel and wealth index. All covariates were measured in a categorical way.

\begin{figure}[H]
\centering 
\includegraphics[width=12cm]{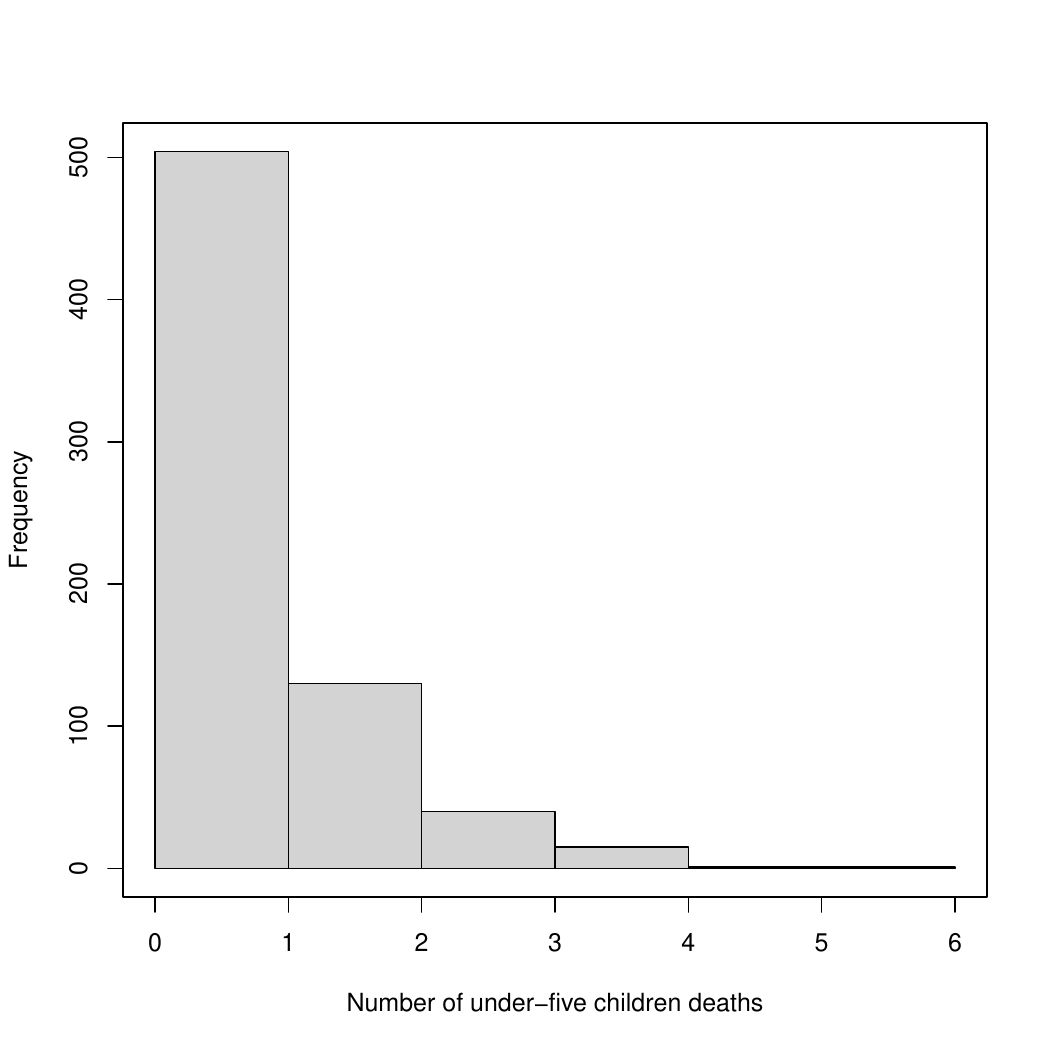} 
\caption{Histogram for the number of under-five children deaths by mother in Oromia}
\label{fig:hist}
\end{figure}

Figure \ref{fig:hist} presents a histogram of the response variable. The number of under-five children deaths by mother in Oromia has an asymmetric distribution and the sample has many zeros ($72.9\%$). This high proportion of zeros suggests that a zero-inflated regression model may be adequate to fit the response variable. Unfortunately, there are mothers in the sample that lost four or five children before they complete five years old.

The ZIP3 regression model was fitted considering a logarithmic link function for $\mu$ and for $\phi$. We selected covariates for the model especially based on the results of likelihood ratio tests. Table \ref{ta:appl_estim} presents the parameter estimates, standard errors and $p$-values of the likelihood ratio tests for the final ZIP3 regression model. Note that the estimates of the parameters associated with mother's age and residence are positive. Therefore, the mean of under-five deaths by mother is higher for older women and for those that live in rural areas. On the other hand, the mean of under-five deaths by mother is lower for women who had more years of formal education.

For a better interpretation of the results of the final ZIP3 regression model, the exponential of the parameter estimates (last column of Table \ref{ta:appl_estim}) were calculated. For example, it is estimated that the mean of under-five deaths by mother is $54.77\%$ lower for mothers that studied at least at secondary level than for those that did not have formal education. This finding suggests that an increase in the investments in formal education can reduce the under-five mortality rate in Oromia. The other parameter estimates can be interpreted in a similar way.

 \begin{table}[H]   
 \caption {The final ZIP3 model
for the under-five mortality in Oromia - Ethiopia.} 
     \vspace{-0.5cm} 
   \begin{center}
\tabcolsep=0.13cm    
   \scalebox{0.99}{
\begin{tabular} {cccrrrr}      
\hline									
Submodel	&	Covariate	&	Category & Estimate	&	Std. Error	& $p$-value &
Exp(estim) \\
\hline									
$\mu$	&	Intercept	&	& $-$2.5390	&	0.3915	&	$< 0.0001$ & 0.0789	\\
\cline{2-7}
 & Mother's age & 15-24 (ref) &  &  &
  &  \\
  &  & 25-34 & 0.9798 & 0.2559 &
 $< 0.0001$ & 2.6638 \\
  & & 35-49 & 1.6787 & 0.2626 & &
  5.3584 \\
  \cline{2-7}
 & Education level & No educ (ref) &  &  &
  &  \\
  &  & Primary & $-$0.5303 & 0.1689 &
 0.0016 & 0.5884 \\
  & & Sec/Higher & $-$0.7934 & 0.4262 & &
  0.4523 \\  
    \cline{2-7}
  & Residence & Urban (ref) &  &  & \multirow{2}*{0.0038} &  \\ 
  & & Rural & 0.7759 & 0.3010 & & 2.1726 \\
\hline									
$\phi$	&	Intercept	& &	$-$1.7370	&	0.3703	&	$< 0.0001$ & 0.1761	\\
\hline
\end{tabular}
}
\end{center}
    \label{ta:appl_estim}
\end{table}

\begin{figure}[H]
     \begin{subfigure}[b]{0.45\textwidth}
         \centering
         \includegraphics[width=\linewidth]{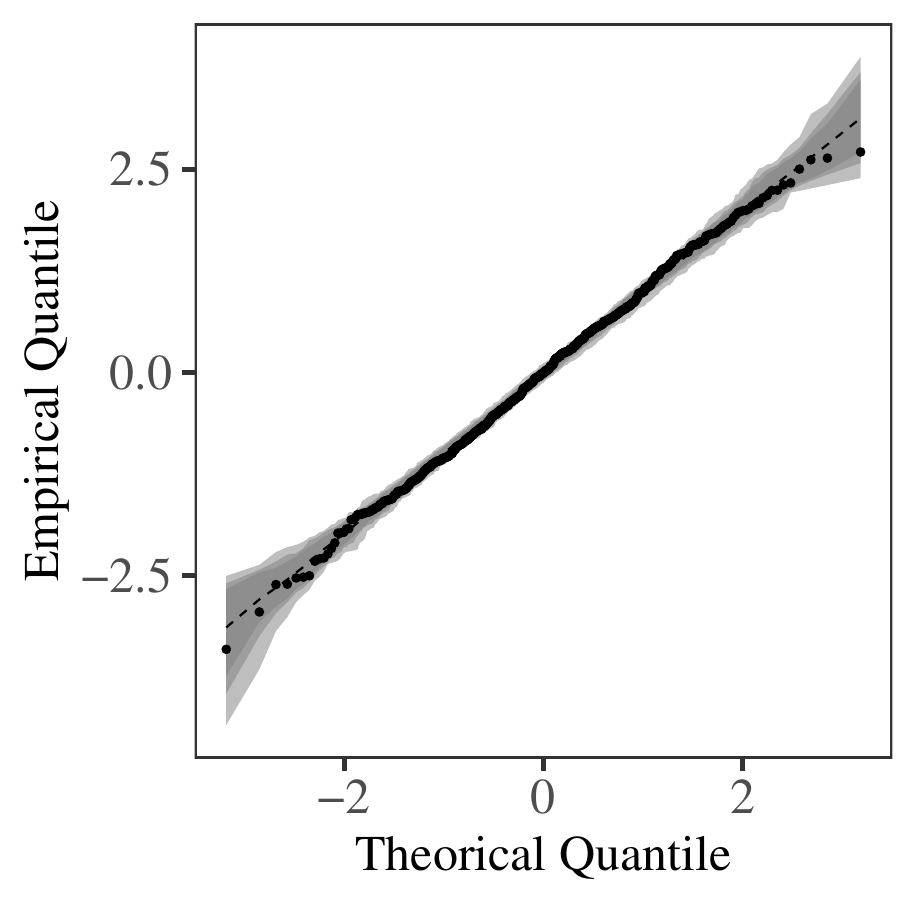}
         \caption{Envelope}
         \label{fig: Logrestic Regression}
     \end{subfigure}
     \hfill
     \begin{subfigure}[b]{0.45\textwidth}
         \centering
         \includegraphics[width=\linewidth]{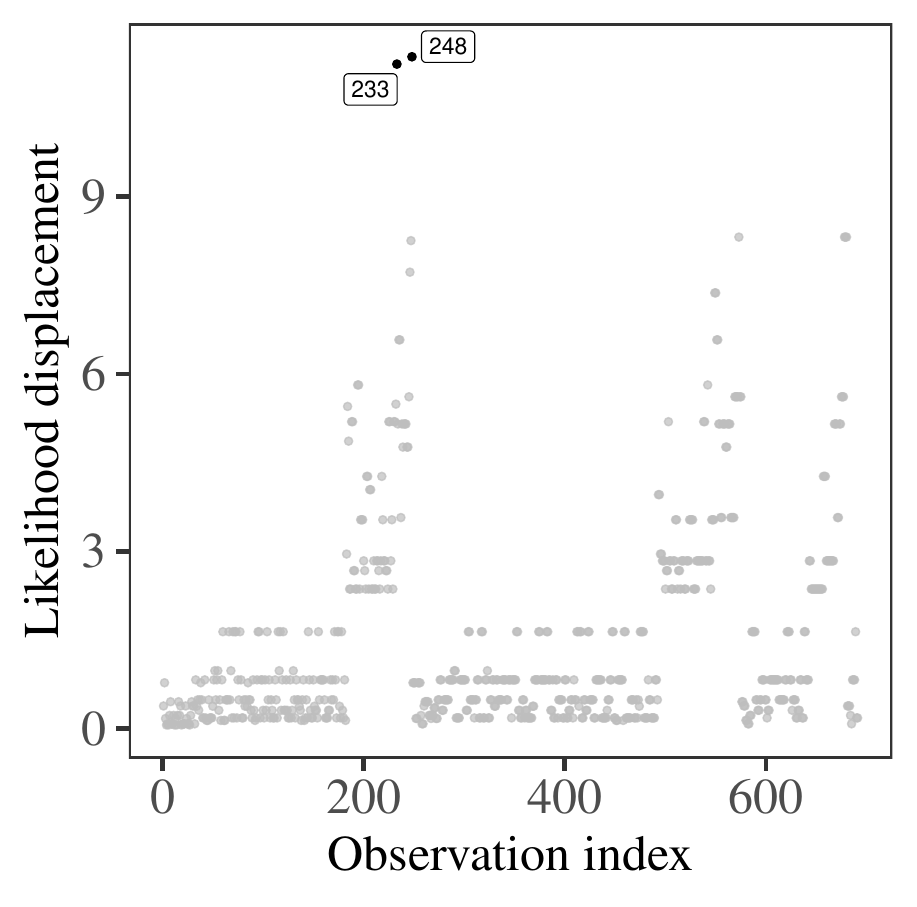}
         \caption{Likelihood displacement}
         \label{fig: Naive Bayes}
     \end{subfigure}
        \caption{Normal probability plot with simulated envelope and index plot of the likelihood displacement for the Oromia - Ethiopia data.}
        \label{fig:diag}
\end{figure}

We used the tools discussed in Section \ref{sec:diagn} to conduct the diagnostic analysis in the final ZIP3 regression model. The left plot of Figure \ref{fig:diag} presents a normal probability plot with simulated envelope \citep{atkinson1981two} using the randomized quantile residual. The plot does not suggest model misspecification. We also obtained the likelihood displacement for the 691 observations (right plot of Figure \ref{fig:diag}). Observations \{233\} and \{248\} have considerably higher values of the likelihood displacement. To study the impact on model inference after removing cases identified as potentially influential, we fitted the model without each of these observations and also without both of them.

Table~\ref{ta:appl_rcs} presents the relative changes (RC) in the parameter estimates and their corresponding changes in the estimated standard errors (RCSE), based on the under-five mortality data. These changes are calculated from
$$
\text{RC}(\hat{\theta}_j)_{(i)} = \left|\frac{\hat{\theta}_j - \hat{\theta}_{j(i)}}{\hat{\theta}_j} \right| \times 100\% \quad \text{and} \quad \text{RCSE}(\hat{\theta}_j)_{(i)} = \left|\frac{\text{SE}(\hat{\theta}_j) - \text{SE}(\hat{\theta}_{j})_{(i)}}{\text{SE}(\hat{\theta}_j) } \right| \times 100\%,
$$
where \(\hat{\theta}_{j(i)}\) and \(\text{SE}(\hat{\theta}_{j})_{(i)}\) represent the ML estimates of  \(j\)th parameter of the model and the estimates of the standard error of the corresponding estimator, respectively, obtained after removing the \(i\)th observation. Note that all RC and RCSE in the three fitted models without one or two potentially influential observations are lower than $14\%$. Note also that the $p$-values of the likelihood ratio tests remain below $5\%$ in these three fitted model. Therefore, the exclusion of these cases do not substantially change the fitted model.

 To investigate if other simple count regression model provides a better fit to the number of under-five children deaths by mother in Oromia than the ZIP3, we fitted three other regression models considering the same covariates of the final model presented in Table \ref{ta:appl_estim}. The considered regression models assume the following distributions for the response variable: Poisson (PO), negative binomial (NB) and zero-inflated negative binomial (ZINB). When we ran the  ZINB regression model considering the same covariates of the final ZIP3 regression model, we obtained an error in the estimation algorithm of the gamlss package. To consider the ZINB regression model in our comparison, we also fitted the three regression models using two out of three selected covariates. Table \ref{ta:aic_bic} presents the AIC and BIC of the four regression models for different choices of covariates. Note that the ZIP3 regression model has the lowest AIC and BIC in all cases, suggesting that this model provides a better fit to the number of under-five children deaths by mother in Oromia than its competitors.

 \begin{table}[H]   
 \caption {RCs (in \%) in ML estimates and in the corresponding estimated standard errors for the indicated removed case(s), and respective \(p\)-values using data on under-five mortality in Oromia - Ethiopia.} 
     \vspace{-0.5cm} 
   \begin{center}
\begin{tabular} {cccrrrr}      
\hline									
Remove cases &Submodel	&	Covariate	&	\multicolumn{1}{c}{Category} & \multicolumn{1}{c}{\(\text{RC}(\hat{\theta})\)} &	\multicolumn{1}{c}{\(\text{RCSE}(\hat{\theta})\)}& \multicolumn{1}{c}{$p$-value}  \\
\hline									
\multirow{7}*{None}&$\mu$	&	Intercept	&	& \multicolumn{1}{c}{\(\times\)}	&	\multicolumn{1}{c}{\(\times\)}	&	$< 0.0001$ \\ \cline{3-7}
& & Mother's age & 25-34 & \multicolumn{1}{c}{\(\times\)}	 & \multicolumn{1}{c}{\(\times\)}	 &
 \multirow{2}*{$< 0.0001$}\\
 & & & 35-49 & \multicolumn{1}{c}{\(\times\)}	 & \multicolumn{1}{c}{\(\times\)}	 & \\ \cline{3-7}
&  & Education level & Primary & \multicolumn{1}{c}{\(\times\)}	 & \multicolumn{1}{c}{\(\times\)}	 & \multirow{2}*{0.0016}  \\ 
&  & & Sec/Higher & \multicolumn{1}{c}{\(\times\)}	 & \multicolumn{1}{c}{\(\times\)}	 &  \\
\cline{3-7}  
&  & Residence & Rural & \multicolumn{1}{c}{\(\times\)}	 & \multicolumn{1}{c}{\(\times\)}	   & 0.0038 \\
\cline{2-7}								
&$\phi$	&	Intercept	& &	\multicolumn{1}{c}{\(\times\)}	&	\multicolumn{1}{c}{\(\times\)}	&	$< 0.0001$ \\
\hline \hline
\multirow{7}*{\{233\}}&$\mu$	&	Intercept	&	& \multicolumn{1}{c}{6.55} &	\multicolumn{1}{c}{3.16} &	$< 0.0001$ \\ \cline{3-7}
& & Mother's age & 25-34 & \multicolumn{1}{c}{6.96} & \multicolumn{1}{c}{2.35} &
\multirow{2}*{$< 0.0001$}\\
 & & & 35-49 & \multicolumn{1}{c}{4.16} & \multicolumn{1}{c}{2.20} & \\ \cline{3-7}
&  & Education level & Primary & \multicolumn{1}{c}{1.92} & \multicolumn{1}{c}{0.31} & \multirow{2}*{0.0014}  \\ 
&  & & Sec/Higher & \multicolumn{1}{c}{2.86} & \multicolumn{1}{c}{0.10} &  \\
\cline{3-7}  
&  & Residence & Rural & \multicolumn{1}{c}{13.28} & \multicolumn{1}{c}{3.97}   & 0.0059 \\
\cline{2-7}								
&$\phi$	&	Intercept	& &	\multicolumn{1}{c}{6.03} &	\multicolumn{1}{c}{11.50}	&	$< 0.0001$ \\
\hline \hline
\multirow{7}*{\{248\}}&$\mu$	&	Intercept	&	& \multicolumn{1}{c}{0.21}	&	\multicolumn{1}{c}{0.17}	&	$< 0.0001$ \\ \cline{3-7}
& & Mother's age & 25-34 & \multicolumn{1}{c}{0.74} & \multicolumn{1}{c}{0.06} &
\multirow{2}*{$< 0.0001$}\\
 & & & 35-49 & \multicolumn{1}{c}{1.47} & \multicolumn{1}{c}{0.18} & \\ \cline{3-7}
&  & Education level & Primary & \multicolumn{1}{c}{2.12} & \multicolumn{1}{c}{0.01} & \multirow{2}*{0.0020}  \\ 
&  & & Sec/Higher & \multicolumn{1}{c}{0.27} & \multicolumn{1}{c}{0.01} &  \\
\cline{3-7}  
&  & Residence & Rural & \multicolumn{1}{c}{2.24} & \multicolumn{1}{c}{0.24}   & 0.0182 \\
\cline{2-7}								
&$\phi$	&	Intercept	& &	\multicolumn{1}{c}{3.96}	&	\multicolumn{1}{c}{5.42}	&	$< 0.0001$ \\
\hline \hline
\multirow{7}*{\{233, 248\}}&$\mu$	&	Intercept	&	& \multicolumn{1}{c}{6.35}	&	\multicolumn{1}{c}{0.17}	&	$< 0.0001$ \\ \cline{3-7}
& & Mother's age & 25-34 & \multicolumn{1}{c}{7.77} & \multicolumn{1}{c}{0.06} &
 \multirow{2}*{$< 0.0001$}\\
 & & & 35-49 & \multicolumn{1}{c}{2.71} & \multicolumn{1}{c}{0.18} & \\ \cline{3-7}
&  & Education level & Primary & \multicolumn{1}{c}{0.23} & \multicolumn{1}{c}{0.01} & \multirow{2}*{0.0018}  \\ 
&  & & Sec/Higher & \multicolumn{1}{c}{3.12} & \multicolumn{1}{c}{0.01} &  \\
\cline{3-7}  
&  & Residence & Rural & \multicolumn{1}{c}{11.03} & \multicolumn{1}{c}{0.24}  & 0.0071 \\
\cline{2-7}								
&$\phi$	&	Intercept	& &	\multicolumn{1}{c}{10.99}&	\multicolumn{1}{c}{5.42}	&	$< 0.0001$ \\
\hline
\end{tabular}
\end{center}
    \label{ta:appl_rcs}
\end{table}

 \begin{table}[H]   
 \caption {AIC and BIC for the four considered regression models.} 
     \vspace{-0.5cm} 
   \begin{center}
\begin{tabular} {crrrrrrrrr}      
\hline		
Covariates in & \multicolumn{4}{c}{AIC} & & \multicolumn{4}{c}{BIC} \\
\cline{2-5} \cline{7-10}
the model$^\star$	&	ZIP3	&	PO	&	NB	&	ZINB	& &	ZIP3	&	PO	&	NB	&	ZINB	\\
\hline
$1,2,3$	&	1033.5	&	1041.1	&	1035.4	&	\multicolumn{1}{c}{\(\times\)} & &	1065.2	&	1068.3	&	1067.1	&	\multicolumn{1}{c}{\(\times\)} \\
$1,2$	&	1039.9	&	1047.3	&	1040.3	&	1072.8	& &	1067.1	&	1070.0	&	1067.5	&	1104.6	\\
$1,3$	&	1042.4	&	1052.9	&	1045.7	&	1047.7	& &	1065.1	&	1071.0	&	1068.4	&	1074.9	\\
$2,3$	&	1083.7	&	1103.0	&	1085.3	&	1087.3	& &	1106.4	&	1121.1	&	1107.9	&	1114.5	\\
\hline
\multicolumn{10}{l}{$^\star$ \small{Covariate 1: mother's age, covariate 2: education level, covariate 3: residence}} 
\end{tabular}
\end{center}
    \label{ta:aic_bic}
\end{table}

\section{Concluding remarks}
\label{sec:conc}
   

In this work, we introduced a more interpretable regression model for count data with excess of zeros based on a reparameterization of the zero-inflated Poisson distribution. Inferential and diagnostic tools for this novel model were discussed. An application on under-five mortality in Oromia, Ethiopia illustrated the usefulness of the proposed regression model. 

The existing zero-inflated regression models are less interpretable than the other common  models for count data. The parameters of our regression model are easily interpreted, especially when using the logarithmic link function as it was done in Section \ref{sec:appl}. Therefore, the proposed ZIP3 regression model will be very useful in medical research and also in other areas, when the response is a count variable with high proportion of zeros.

\section{Acknowledgment}

This work was partially supported by São Paulo Research Foundation (FAPESP), grant number 2020/16334-9.

\singlespacing   

\bibliographystyle{elsarticle-harv}

\bibliography{bibzip}  

\end{document}